\documentclass[twocolumn,showpacs,nofootinbib]{revtex4}%
\usepackage{amssymb}
\usepackage{graphicx}%
\usepackage{amsmath}%
\usepackage{amsfonts}
\begin{document}
\title{Deposit Growth in the Wetting of an Angular Region with Uniform Evaporation}
\author{Rui Zheng}
\email{ruizheng@uchicago.edu}
\author{Yuri O. Popov}
\altaffiliation{Current address: Department of Physics, University of Michigan, 450 Church
St., Ann Arbor, Michigan 48109, USA}
\author{Thomas A. Witten}
\affiliation{James Franck Institute and Department of Physics, University of Chicago, 5640
S. Ellis Ave., Chicago, Illinois 60637, USA}

\date{\today }

\begin{abstract}
Solvent loss due to evaporation in a drying drop can drive capillary flows and
solute migration. The flow is controlled by the evaporation profile and the
geometry of the drop. We predict the flow and solute migration near a sharp
corner of the perimeter under the conditions of uniform evaporation. This
extends the study of Ref.\ 6, which considered a singular evaporation profile,
characteristic of a dry surrounding surface. We find the rate of the deposit
growth along contact lines in early and intermediate time regimes. Compared to
the dry-surface evaporation profile of Ref.\ 6, uniform evaporation yields
more singular deposition in the early time regime, and nearly uniform
deposition profile is obtained for a wide range of opening angle in the
intermediate time regime. Uniform evaporation also shows a more pronounced
contrast between acute opening angles and obtuse opening angles.

\end{abstract}
\pacs{47.55.Dz, 68.03.Fg, 81.15.-z}
\maketitle

\section{Introduction}

Evaporative contact line deposition, the \textquotedblleft coffee-drop
effect\textquotedblright, has been the subject of several recent
papers\ \cite{1, 2, 3, 4, 5, 6, 7, -8}. The physical problem originates from a
simple phenomenon of everyday life: when a drop containing a solute such as
coffee dries on a surface, the solute is driven to the contact line, forming a
characteristic ring pattern. This simple phenomenon is potentially important
in many areas of both scientific and industrial applications \cite{-1, -2, -3,
-4}. This evaporation mechanism can create very fine lines of deposition in a
robust way that requires no explicit forming. Further, it is a way of
concentrating material strongly in a quantitatively predictable way. Lastly,
it creates capillary flow patterns that can be useful for processing of
polyatomic solutes like DNA \cite{-5, -6, -7}.

\begin{figure}
[ptb]
\begin{center}
\includegraphics[
natwidth=10.0024in, 
natheight=7.4936in,
width=6.3236in,
]
{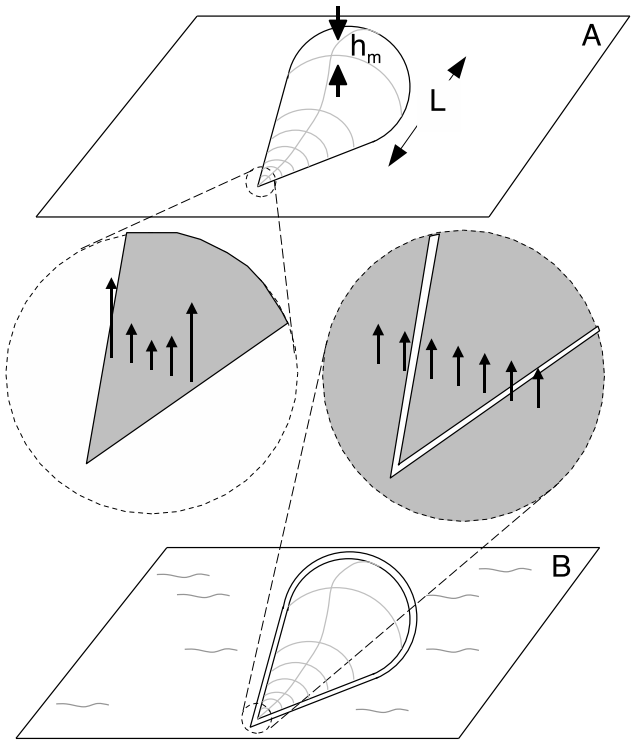}%
\caption{Sketch of the two different evaporation conditions contrasted in the
text. Both pictures show a small liquid drop of size $L$ and maximum thickness
$h_{m}$ with part of its edge constrained to an angular corner shape by means
of \textit{e.g. }two scratches on the surface. Figure A shows a drop on a dry
surface. Magnified tip region indicates the profile of evaporating flux across
the surface by a row of arrows. Shading indicates the wet region. The flux
diverges at the edges. Figure B shows a similar drop surrounded by a wet
surface. Magnified region shows the uniform evaporating flux.}%
\label{AB}%
\end{center}
\end{figure}

One striking aspect of this deposition phenomenon is its dependence on the
shape of the droplet. This dependence was recently explored by Popov and
Witten \cite{5,6}, who studied corner-shaped drops. Here the liquid region on
the surface has the form of a sector of arbitrary opening angle $\alpha$
(Fig.\ \ref{AB}). Such shapes contrast strongly with the circular drops
treated in previous studies \cite{1,2,3,4,-8}. Popov and Witten found that
this difference in shape led to striking differences in evaporative flow and
deposition near the apex of the drop. Both the flow and the deposition profile
showed singular power-law behavior as a function of distance $r$ from the tip.
These power laws vary in a predicted way with the opening angle, but are
otherwise universal. The growth of the deposition also shows several different
predicted behaviors in three defined time regimes. Thus by manipulating the
shape of a droplet one has extensive control over its deposition properties.

In this paper we further explore the range of control possible in evaporative
deposition. This study is motivated by a finding from Ref.\ 6: the power laws
governing the deposition depend on the evaporation conditions in the vicinity
of the drop. Ref.\ 6 considered the usual diffusion-controlled evaporation
conditions, in which the liquid is surrounded by a dry surface and the
evaporation rate is limited by the diffusion of vapor away from the drop. In
these conditions concentration $n(r)$ in the air obeys Laplace's equation with
$n$ at the surface set by the fixed saturation concentration. As in the
analogous electrostatic problem, the gradient of $n$ diverges at the edge and
at the tip. The resulting divergent evaporation profiles contribute strongly
to the controllable deposition properties found in Ref.\ 6.

One may readily alter these evaporation conditions, and strong differences in
the deposition are expected to result. We consider a condition that contrasts
strongly with the singular evaporation treated in Ref.\ 6, \textit{viz}.
\textit{uniform} evaporation. This contrast is illustrated in Fig.\ \ref{AB}.
Uniform evaporation can be created by surrounding the drop by a wet surface
instead of a dry one, as illustrated in Figure 1B. The wet-surface evaporation
case allows us to discern the role of the evaporation profile in producing the
power-law behaviors revealed by Ref.\ 6. The wet-surface evaporation case is
also mathematically simpler to treat than the dry-surface case of Ref.\ 6.
This allows us to make cleaner predictions with fewer approximations for the
wet-surface case.%

Our work is to a large extent a mere application of the general theory worked
out in Ref.\ 6. We shall follow the same approach and exposition for the
current case that was used in Ref.\ 6. Before working out the quantitative
behavior we describe the experiment qualitatively. A droplet of linear
dimension $L$ is forced to have a corner shape over part of its perimeter. One
may fix the edge of the drop \textit{e.g.} by making shallow scratches in the
surface. These serve to pin the contact line at the scratch. We choose $L$
smaller than a few millimeters so that gravitational effects on the droplet's
shape are minor. We denote the maximal thickness of the drop by $h_{m}$ and
keep the droplet volume small enough to assure that $h_{m}$ $\ll$ $L$. Once
the evaporation starts, the volume diminishes at a constant rate, and thus
$h_{m}$ decreases linearly with time. At some final time $t_{f}$ this
thickness extrapolates to zero. Like Reference 6, we restrict our attention to
time much less than this $t_{f}$, and it will show later that for $t\ll t_{f}%
$, in the early drying stages, the time dependence of $L$ and $h_{m}$ can be
actually ignored. Our predictive power is strongest for this regime.

The shape of the droplet during this thinning process is dictated by its
surface tension. Near the edges, the local reduction in volume owing to
thinning is much smaller than the local loss of volume to evaporation. Thus a
flow towards the edge is needed in order to replace the evaporative loss. Any
solute suspended in the fluid is carried along by this flow. The asymptotic
flow near the tip is minimally influenced by the bulk of the drop, and thus
this flow can be readily calculated. Knowing this flow profile, one may deduce
how much solute should be carried to a given point on the edge in a given time
$t$. This amount grows in a characteristic way with time and with distance
from the tip. Our aim is to see how the deposition is influenced by opening
angle and how much this deposition differs from the dry-surface results of
Ref.\ 6. The main difference is in the opposite direction from what one might
expect. We find that the wet-surface evaporation leads to deposition that is
more concentrated at the tip than the dry-surface deposition of Ref.\ 6. This
is despite the diverging evaporation at the edges and tip produced in the
dry-surface case.

The paper is organized in parallel with Ref.\ 6. First, we review the basic
physical model and its mathematical framework, which was introduced in
Ref.\ 6. Next, the system is solved both analytically and numerically: the
flow field is described, and its asymptotic properties studied. We obtain the
power law of the deposition rate in early time regime and intermediate time
regime. Then we compare our results with those of the dry-surface case. More
discussion and conclusions follow in the last sections.

\section{Principles governing the deposition}

\begin{figure}
[ptb]
\begin{center}
\includegraphics[
height=3.2854in,
width=4.12in
]%
{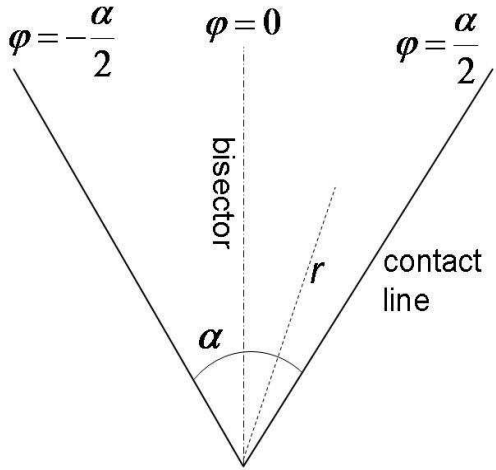}%
\caption{Sketch of the angular drop showing geometrical quantities used in the
text}%
\label{figsetting}%
\end{center}
\end{figure}

Following Ref.\ 5, we consider a droplet of very dilute suspension bounded in
an angular region of opening angle $\alpha$, as illustrated in Fig.\ \ref{AB}.
We use cylindrical coordinates $(r$, $\varphi$, $z)$ defined in
Fig.\ \ref{figsetting} with the range $0<r<\infty$ and $-\alpha/2<\varphi
<\alpha/2$, to describe our system. $z$ is the coordinate normal to the
substrate. We first present the governing equations in generality and then
specialize to the regime of interest: thin drops undergoing slow flows.%

The equilibrium shape of the surface is dictated by the minimization of
surface energy. The minimum energy surface has uniform mean curvature $H$.
Specifically, if the surface tension of the liquid is $\sigma$ and the
pressure difference across the surface is $\Delta p$, then the balance of
normal forces dictates:%
\begin{equation}
\Delta p=-2H\sigma.\label{1}%
\end{equation}
In the sequel, we shall represent the surface by its thickness profile $h(r$,
$\varphi$, $t)$. The mean curvature $H$ depends on the local derivatives of
$h$, to be specified below.

As indicated in the Introduction, evaporation induces a flow towards the edge
of the drop. Denoting the local velocity by $\mathbf{u}(r$, $\varphi$, $z$,
$t)$ with in-plane component $\mathbf{u}_{s}$, it is useful to define a
depth-averaged velocity field $\mathbf{v}$
\begin{equation}
\mathbf{v=}\frac{1}{h}\int_{0}^{h}\mathbf{u}_{s}dz.\label{2}%
\end{equation}
The condition for local mass conservation may be stated in terms of this
$\mathbf{v}$:%
\begin{equation}
\nabla\cdot(h\mathbf{v)+}\frac{J_{0}}{\rho}\sqrt{1+(\nabla h)^{2}}%
+\partial_{t}h=0.\label{3}%
\end{equation}
Here $\rho$ is the density of the fluid, and $J_{0}$ the mass loss per unit
projected area and time at the point in question. In Ref.\ 6, this $J_{0}$ is
a strong function of position, and it diverges at the contact line. In the
present work, it is a mere constant. By itself, this condition is not
sufficient to determine $\mathbf{v}$. Ultimately, $\mathbf{v}$ is determined
by Newton's equations on each fluid element. We shall consider the creeping
flow regime where forces are in near equilibrium and acceleration plays a
negligible role in Newton's equations. Then Newton's equations reduce to the
Stokes equation:
\begin{equation}
\nabla p=\eta\nabla^{2}\mathbf{u,}\label{4}%
\end{equation}
where $p$ is the fluid pressure, and $\eta$ is the dynamic viscosity. The lack
of inertia implies that $\mathbf{v}$ is a potential-like flow, as shown below.

Further physical considerations and simplifications are needed to solve the
system analytically. Firstly, as we are considering a thin drop, there is a
separation of the vertical and horizontal scales in this problem, and several
simplifications follow \cite{6}. The pressure inside the drop $p$ does not
depend on the $z$ coordinate $\partial_{z}p=0$. The surface of the drop should
have a small slope $\left\vert \nabla h\right\vert \ll1$. And the
$z$-derivatives of flow dominate, i.e., $\left\vert \partial_{z}%
u_{i}\right\vert \gg\left\vert \partial_{s}u_{i}\right\vert $, where $s$
represents any direction parallel to the substrate plane, and $u_{i}$ refers
to any velocity component. Under these considerations, (\ref{4}) has the form
\cite{6}%
\begin{equation}
\nabla_{s}p=\eta\partial_{zz}\mathbf{u}_{s}.\label{5}%
\end{equation}
With boundary conditions: $\mathbf{u}_{s}|_{z=0}=0,\partial_{z}\mathbf{u}%
_{s}|_{z=h}=0$, we obtain
\begin{equation}
\mathbf{u}_{s}=\frac{\nabla p}{\eta}\left(  \frac{z^{2}}{2}-hz\right)
.\label{6}%
\end{equation}
With expressions (\ref{2}) and (\ref{6}), we have \cite{6}
\begin{equation}
\mathbf{v=-}\frac{h^{2}}{3\eta}\nabla p.\label{7}%
\end{equation}
Thus $\mathbf{v/}h^{2}$ can be represented as a gradient of the scalar
potential as announced above. The flow in the drop clearly depends on the
relative importance of surface forces and viscous forces. This importance is
ordinarily characterized by the capillary number $Ca=v\eta/\sigma$. For
water-like fluids, the capillary number is small whenever $v\ll100$ $%
\operatorname{m}%
/%
\operatorname{s}%
$. Ordinary evaporating flows are in the range of $10^{-5}$ $%
\operatorname{m}%
/%
\operatorname{s}%
$; accordingly, viscous forces may be considered as weak in comparison to
surface tension. Thus we anticipate that the shape of the drop is nearly the
equilibrium shape in the absence of the flow. To establish this formally and
systematically, we express the pressure and height as expansions in the
capillary number: $p=p_{0}+(Ca)p_{1}+(Ca)^{2}p_{2}+...$ and $h=h_{0}%
+(Ca)h_{1}+(Ca)^{2}h_{2}+...$. By using these expansions in Eqs.\ (\ref{1}),
(\ref{3}) and (\ref{7}), we find the lowest order results \cite{6}
\begin{equation}
2H=-\frac{p_{0}-p_{atm}}{\sigma},\label{8}%
\end{equation}
\begin{equation}
\nabla\cdot(h_{0}^{3}\nabla\psi)=-\frac{J_{0}}{\rho}-\partial_{t}%
h_{0},\label{9}%
\end{equation}
\begin{equation}
\mathbf{v}_{0}\mathbf{=}h_{0}^{2}\nabla\psi,\label{10}%
\end{equation}
where $\psi=-p_{1}(Ca)/(3\eta),$ and the leading term $p_{0}$ does not vary
with local coordinates and is only a function of time $t$. Thus, one can use
Eq.\ (\ref{8}) to determine the equilibrium drop surface shape $h_{0}$, then
solve Eq.\ (\ref{9}) with respect to $\psi(r,\varphi,t)$, and finally
determine the velocity field from Eq.\ (\ref{10}). To simplify the notations,
we will write $p_{0}-p_{atm}$ as $\triangle p$, and $h_{0}$ as $h$ in the rest
of the paper.

Eq.\ (\ref{8}) introduces a length scale into the problem, i.e., the mean
radius of curvature $R(t)=\sigma/\Delta p$. Since the evaporation rate is
constant in time, the droplet volume and thus its thickness decrease linearly
in time until the terminal time $t_{f}$. This means \cite{6} that the mean
curvature $R(t)\propto L^{2}/h_{m}$ can be written as
\begin{equation}
R(t)=\frac{R_{i}}{1-t/t_{f}},\label{10d}%
\end{equation}
where $R_{i}$ is the initial mean radius of curvature, and $t_{f}$ is the
total drying time. Thus in the early drying stages ($t\ll t_{f}$), which is
the only case we are to consider in this paper, the time dependence of $R$ can
be ignored, and the shape of the drop can be assumed to vary with time
adiabatically (i.e., slowly compared to all other processes). We may generally
treat $R(t)$ as constant $R_{i}$ in the rest of the paper, keeping in mind its
time dependence (\ref{10d}).

\section{Behavior of the drop near the tip}

\subsection{Surface shape}

\begin{figure}
[ptb]
\begin{center}
\includegraphics[
natheight=3.338200in,
natwidth=4.122600in,
height=3.0926in,
width=3.4748in
]%
{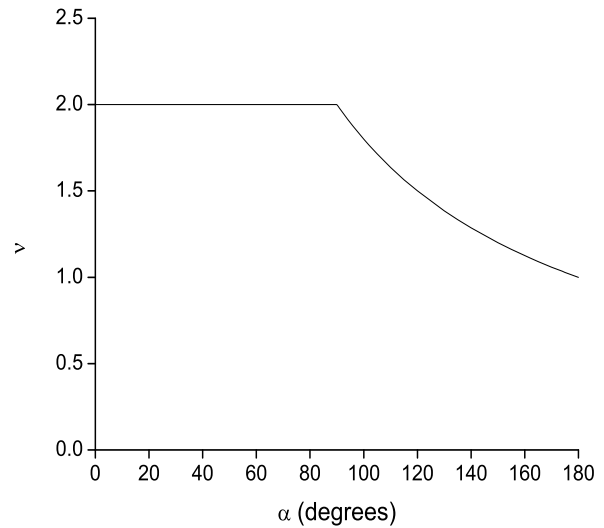}%
\caption{Relation between the exponent $\nu$ and the opening angle $\alpha$}%
\label{nu}%
\end{center}
\end{figure}

If written explicitly in terms of $h(r,\varphi,t)$, equation (\ref{8}) reduces
to the Poisson equation under the assumption of small slope of the surface:
\begin{equation}
\nabla^{2}h=-\frac{\Delta p}{\sigma},\label{11}%
\end{equation}
with boundary conditions $h(r,-\alpha/2)=h(r,\alpha/2)=h(0,\varphi)=0$
\cite{5}. Here $\sigma/\Delta p=R(t)\approx R_{i}$ is the mean radius of
curvature, and we are only interested in the asymptotic limit $r\ll R(t)$.
This problem is quite classical \cite{8}, however, its results are somewhat
unexpected \cite{5}. The leading term of the solution in the limit $r\ll R(t)
$ is different for acute and obtuse angles and can be obtained by either
solving the small-slope (horizontal) equation (\ref{11}) or by the series
expansion of the full equation (\ref{8}). By both methods, the lowest order
term in $r$ of the series expansion of the surface shape $h$ was found in
Ref.\ 5:
\begin{equation}
h(r,\varphi,t)=\frac{r^{\nu}}{R(t)^{\nu-1}}\tilde{h}(\varphi),\label{13}%
\end{equation}
where
\begin{align}
\tilde{h}(\varphi)  & =\frac{1}{4}\left(  \frac{\cos2\varphi}{\cos\alpha
}-1\right)  \text{, }\nu=2\text{, \ \ }0\leq\alpha<\frac{\pi}{2},\label{14a}\\
\tilde{h}(\varphi)  & =C\left(  \alpha\right)  \cos\frac{\pi\varphi}{\alpha
}\text{, \ \ \ \ }\nu=\frac{\pi}{\alpha}\text{, \ }\frac{\pi}{2}<\alpha\leq
\pi.\label{14b}%
\end{align}
Exponent $\nu$ as a function of the opening angle $\alpha$ is shown in
Fig.\ \ref{nu}. The acute-angle solution (\ref{13}), (\ref{14a}) is
self-similar and independent of the remote boundary condition specifying the
surface shape in the bulk of the drop. The obtuse-angle solution (\ref{13}),
(\ref{14b}) has a pre-factor $C\left(  \alpha\right)  $, which does depend on
the remote boundary condition in the bulk of the drop \cite{8} and has the
form \cite{5}:
\begin{equation}
C(\alpha)=\frac{1}{4\alpha-2\pi}+C_{0}+O\left(  \alpha-\frac{\pi}{2}\right)
,\label{15}%
\end{equation}
where $C_{0}$ is \textit{independent} of $\alpha$ and is determined by the
remote boundary condition. Clearly, the leading-order solution (\ref{13}),
(\ref{14a}), and (\ref{14b}) diverges when $\alpha$ approaches the right angle
from either side.%

The notable difference between acute and obtuse opening angles in the main
order results (\ref{13}), (\ref{14a}), and (\ref{14b}) was first discussed in
Ref.\ 17. For acute opening angles, a \textit{local similarity} solution is
valid, i.e., the solution that has no arbitrary coefficients and is
independent of the remote boundary conditions in the bulk of the drop that
need to be further prescribed. Physically, the surface shape of the drop near
the tip is controlled locally and is independent of the physical conditions
far from the tip. For obtuse opening angles, however, the local similarity
does not hold, and the surface shape is no longer controlled locally.
Accordingly, the coefficient $C\left( \alpha\right) $ depends on the remote
boundary conditions. This contrast between acute and obtuse opening angles
leads to different flow behavior and different deposition properties as will
be shown later.

The dependence of $C\left( \alpha\right) $ on the remote boundary condition
can be illustrated in the following way. $C\left( \alpha\right) $ can be
expressed in terms of the linear dimension $L$ and the maximal thickness
$h_{m}$, which characterize the global shape of the drop. Along the bisector
$\varphi=0$, let $r\rightarrow L$ and $h\rightarrow h_{m}$, and using
approximation $R(t)\approx R_{i}\propto L^{2}/h_{m}$, one can obtain from
Eqs.\ (\ref{13}) and (\ref{14b}):
\begin{equation}
C\left(  \alpha\right)  \propto\left(  \frac{L}{h_{m}}\right)  ^{\nu
-2}.\label{1501}%
\end{equation}
Clearly, $C\left( \alpha\right) $ is small since $\nu<2$ for obtuse opening
angles, and its time dependence is weak in the early drying stages and can be ignored.

For numerical purposes, $C_{0}$ in (\ref{15}) will be set to unity, as was
done in Ref.\ 6. This is justified by the fact that when $\alpha$ approaches
$\pi/2$ from above, the divergent term $1/(4\alpha-2\pi)$ dominates, and the
properties of the system become independent of $C_{0}$. In the opposite limit,
$\alpha=\pi$, the tip of the angular region can be chosen at any point on the
contact line due to the symmetry, and physical properties of the drop are
again independent of $C_{0}$. Although for arbitrary obtuse opening angle
$C_{0}$ is controlled by the remote boundary conditions, numerical studies
have shown consistent results that did not vary substantially with $C_{0}$.

The special case $\alpha=\pi/2$ invites further explanation. According to
expressions (\ref{14a}) and (\ref{14b}), function $\tilde{h}(\varphi)$
diverges when $\alpha$ approaches $\pi/2$ from either side. This divergence is
artificial, however \cite{5}. This issue is explained in some detail in
Appendix B. In the rest of the work, we will treat $\alpha=\pi/2$ as the
limiting case and will keep in mind the possible divergence of expressions
(\ref{14a}) and (\ref{14b}).

\subsection{Reduced pressure}

Now we are ready to solve Eq.\ (\ref{9}). We are interested only in the
asymptotic behavior when $r\rightarrow0$ (or $r\ll R $), and in this limit
$\partial_{t}h\varpropto r^{\nu}$ as we know from Eq.\ (\ref{13}). Thus,
$\partial_{t}h$ can be safely dropped compared to the constant term
$J_{0}/\rho$ on the right-hand side of Eq.\ (\ref{9}). Physically, in this
asymptotic limit during the early drying stages the fluid mass transport due
to the gradient of the flow flux is uniquely balanced by the evaporation from
the surface locally at each moment, with the mass change brought about by the
local height change being a higher order small quantity that can be ignored.

The asymptotic solution for the reduced pressure $\psi(r,\varphi,t)$ of
Eq.\ (\ref{9}) can be expressed as:%
\begin{equation}
\psi(r,\varphi,t)=\frac{J_{0}}{\rho}\frac{r^{2-3\nu}}{R(t)^{3-3\nu}}%
\tilde{\psi}\left(  \varphi\right)  ,\label{16}%
\end{equation}
where the exponent of $r$ is determined by simply counting the powers of $r$
on the left side of Eq.\ (\ref{9}). From Eqs. (\ref{9}) and (\ref{16}), we
find explicitly the differential equation that $\tilde{\psi}\left(
\varphi\right)  $ should satisfy \cite{6}
\begin{equation}
\frac{d^{2}\tilde{\psi}}{d\varphi^{2}}+\frac{3}{\tilde{h}}\frac{d\tilde{h}%
}{d\varphi}\frac{d\tilde{\psi}}{d\varphi}-2\left(  3\nu-2\right)  \tilde{\psi
}=-\frac{1}{\tilde{h}^{3}}\text{.}\label{17}%
\end{equation}
This equation depends implicitly on the opening angle, as $-\alpha
/2\leq\varphi\leq\alpha/2$, with $\nu$ and $\tilde{h}$ depending on $\alpha$
as shown in Eqs.\ (\ref{14a}) and (\ref{14b}).

The boundary conditions associated with Eq.\ (\ref{17}) need to be clarified.
Firstly, we expect the flow field to be totally symmetrical with respect to
the bisector $\varphi=0$, and therefore $\tilde{\psi}$ should be even in
$\varphi$:
\begin{equation}
\left. \frac{d\tilde{\psi}}{d\varphi}\right\vert _{\varphi= 0} = 0.\label{19}%
\end{equation}
Secondly, the outer boundary condition at $\varphi=\alpha/2$ can be identified
explicitly as
\begin{equation}
\tilde{h}^{3}\left.  \frac{d\tilde{\psi}}{d\varphi}\right\vert _{\varphi=
\alpha/2} = 0.\label{222}%
\end{equation}
In Appendix A, we justify this boundary condition, and discuss how
Eq.\ (\ref{17}) may be regularized.

The boundary problem of Eq.\ (\ref{17}) with boundary conditions (\ref{19})
and (\ref{222}) is complete and has a unique solution. We show the numerical
solutions $\tilde{\psi}\left(  \varphi\right) $ for typical opening angles
$\alpha=\pi/4$ and $3\pi/4$ in Fig.\ \ref{newintegration}. The analytical
solution to this boundary problem and its asymptotics are also discussed in
Appendix A.
\begin{figure}
[ptb]
\begin{center}
\includegraphics[
trim=0.000000in 0.000000in 0.000000in -0.001625in,
natheight=4.062900in,
natwidth=5.235600in,
height=9.2654cm,
width=10.177cm
]%
{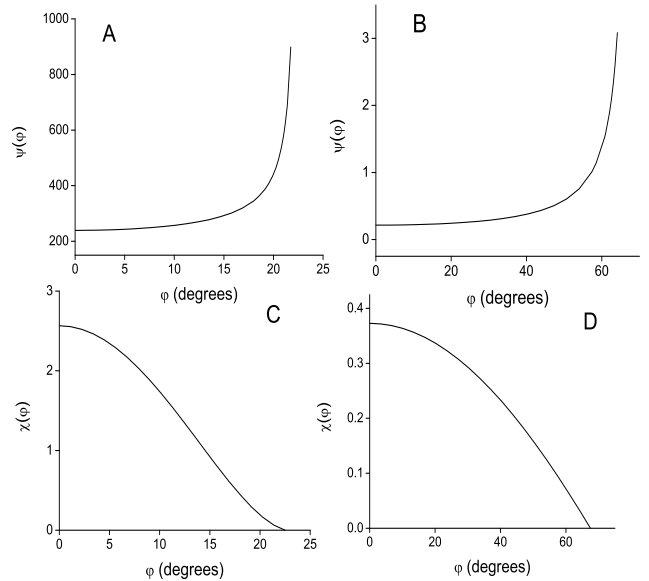}%
\caption{A, B: The reduced pressure function $\tilde{\psi}(\varphi)$ for
representative acute and obtuse opening angles. A: $\alpha=\pi/4$, B:
$\alpha=3\pi/4$; C, D: The regularized function $\tilde{\chi}(\varphi)$,
defined in Appendix A as $\tilde{\chi}(\varphi)=\tilde{h}^{2}\tilde{\psi
}(\varphi)$, corresponding to A and B }%
\label{newintegration}%
\end{center}
\end{figure}

Eq.\ (\ref{17}) is solvable analytically for special opening angle $\alpha
=\pi$. If written in terms of the Cartesian coordinates (where the $x$ axis is
the contact line and the $y$ axis is the bisector), the reduced pressure
function $\psi$ can be found from Eqs. (\ref{14b}), (\ref{16}), and
(\ref{17}), and has the form
\begin{equation}
\psi=\frac{J_{0}}{\rho}\frac{1}{C^{3}(\pi)}\frac{1}{y}.\label{301}%
\end{equation}
Since is does not depend on $x$, this result is fully consistent with the
symmetry of the system when $\alpha=\pi$.

For $0\leq\alpha<\pi$, no special angle exists to reduce the complexity of the
equation. Angle $\pi$ is the only opening angle where we can obtain analytic
results in a closed form. For $\alpha=\pi$ the simplicity is well anticipated,
since this limiting case has no apex at all. Without an apex, all points on
the boundary are equivalent, and most of the resulting properties follow by
symmetry. We will use the exact solution at the opening angle $\pi$ to test
our numerical results and analytical asymptotics.

\section{Results for the physical properties of the system}

\subsection{Flow field}

\begin{figure}
[t]
\begin{center}
\includegraphics[
height=1.727in,
width=3.4394in
]%
{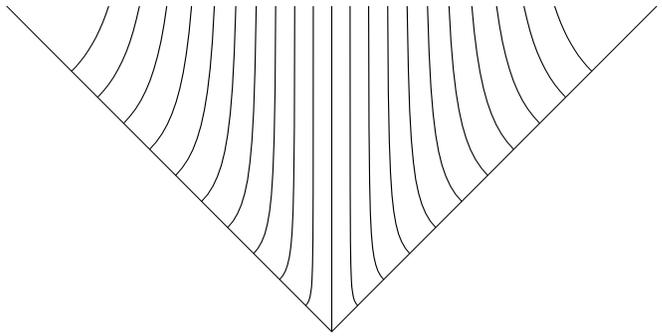}%
\caption{Streamline configuration for the opening angle $\pi/2$}%
\label{flowfield}%
\end{center}
\end{figure}

In polar coordinates, the velocity field (\ref{10}) can be expressed
explicitly:%
\begin{equation}
\mathbf{v=}v_{r}\hat{r}+v_{\varphi}\hat{\varphi},\label{33}%
\end{equation}%
\begin{align}
v_{r}  & =-(3\nu-2)\frac{J_{0}}{\rho}\left(  \frac{r}{R(t)}\right)  ^{1-\nu
}\tilde{h}^{2}\tilde{\psi},\label{34}\\
v_{\varphi}  & =\frac{J_{0}}{\rho}\left(  \frac{r}{R(t)}\right)  ^{1-\nu
}\tilde{h}^{2}\frac{d\tilde{\psi}}{d\varphi}.\label{35}%
\end{align}
As we assume that solute particles carried by the fluid move with the same
velocity as the flow itself (this assumption was actually confirmed both
theoretically and experimentally, and a theoretical estimate can be found in
\cite{-9}), the trajectory of each particle is identified as the streamline of
the flow field, and can be obtained by integrating the velocity field
\cite{6}:
\begin{equation}
\frac{dr}{rd\varphi}=\frac{v_{r}}{v_{\varphi}}=-(3\nu-2)\tilde{\psi}\left(
\frac{d\tilde{\psi}}{d\varphi}\right)  ^{-1}.\label{36}%
\end{equation}
If the particle eventually arrives at $\left(  r_{0},\alpha/2\right)  $ on the
contact line, the trajectory reads as \cite{6}%
\begin{equation}
r(\varphi)=r_{0}\exp\left(  \left(  3\nu-2\right)  \int_{\varphi}^{\alpha
/2}\tilde{\psi}\left(  \frac{d\tilde{\psi}}{d\varphi^{\prime}}\right)
^{-1}d\varphi^{\prime}\right)  .\label{37}%
\end{equation}
In Fig.\ \ref{flowfield} we show the flow field configuration for the opening
angle $\alpha=\pi/2$ computed numerically with the acute-angle expression
(\ref{14a}).

These flow trajectories are independent of the dimensional quantities like the
evaporation rate $J$, and they depend only on the opening angle \cite{6}.
Thus, the trajectory of the moving solute particle does not depend on how fast
the drop evaporates and how thin the liquid layer is.

We now consider the asymptotic properties of the streamlines. For
$\varphi\rightarrow\alpha/2$, velocity component $v_{r}$ vanishes due to the
outer boundary condition, which amounts to the statement that no mass flows
inward or outward along the contact line, while component $v_{\varphi}$
remains finite due to the asymptotic behavior of $d\tilde{\psi}/d\varphi$
(A11), and goes as
\begin{equation}
v_{\varphi}\rightarrow\frac{J_{0}}{\rho}\left(  \frac{r}{R(t)}\right)
^{1-\nu}\left\vert \frac{d\tilde{h}}{d\varphi}\left(  \frac{\alpha}{2}\right)
\right\vert ^{-1},\label{100}%
\end{equation}
or, if written explicitly by employing results (\ref{14a}) and (\ref{14b}),
\begin{align}
v_{\varphi}  & \rightarrow\frac{2J_{0}}{\rho}\frac{1}{\tan\alpha}\left(
\frac{r}{R(t)}\right)  ^{-1}\qquad\ 0\leq\alpha<\frac{\pi}{2},\nonumber\\
v_{\varphi}  & \rightarrow\frac{J_{0}}{\rho}\frac{\alpha}{\pi C(\alpha
)}\left(  \frac{r}{R(t)}\right)  ^{1-\pi/\alpha}\text{ \ }\frac{\pi}{2}%
<\alpha\leq\pi.\label{101}%
\end{align}
In particular, $v_{\varphi}$ seemingly vanishes as $\alpha\rightarrow\pi/2$.
However, in fact it does not. We explore this point in Appendix B.
Consequently, as $\varphi\rightarrow\alpha/2$, $dr/d\varphi\rightarrow0$, and
the streamlines are perpendicular to the contact line.

In the opposite limit, $\varphi\rightarrow0$, the streamlines reach far into
the bulk of the drop, and function $r\left(  \varphi\right)  $ is
divergent.\footnote{Since $\mathbf{v}/h^{2}$ is a potential flow, the
streamlines $r(\varphi)$ may end only where the reduced pressure $\psi$
assumes local extreme values, i.e., at the contact line and at infinity (in
the bulk of the drop).} Behavior of the streamlines depends on the asymptotic
properties of $\tilde{\psi}$ near the bisector, which, according to
Eq.\ (\ref{17}) and boundary condition (\ref{19}), can be written as
\begin{equation}
\tilde{\psi}\rightarrow\tilde{\psi}(0)+\frac{1}{2}\left(  2(3\nu-2)\tilde
{\psi}(0)-\frac{1}{\tilde{h}^{3}(0)}\right)  \varphi^{2}.\label{38}%
\end{equation}
The boundary value $\tilde{\psi}(0)$ requires the complete solution of
Eq.\ (\ref{17}). The divergent part of the integration in (\ref{37}) is then
uniquely determined by the asymptotic form (\ref{38}) since (Eq.\ (50) in
Ref.\ 6):
\begin{equation}
\int_{\varphi}^{\alpha/2}\tilde{\psi}\left(  \frac{d\tilde{\psi}}%
{d\varphi^{\prime}}\right)  ^{-1}d\varphi^{\prime}\rightarrow\frac{1}%
{2(3\nu-2)-\kappa^{2}}\ln\frac{\alpha}{2\varphi},\label{39}%
\end{equation}
where
\begin{equation}
\kappa^{2}=\frac{1}{\tilde{h}^{3}(0)\tilde{\psi}(0)}.\label{40}%
\end{equation}
Therefore the streamlines in this limit scale as
\begin{equation}
r\rightarrow r_{0}\left(  \frac{\alpha}{2\varphi}\right)  ^{\epsilon}%
,\qquad\varphi\rightarrow0,\label{41}%
\end{equation}
with%
\begin{equation}
\epsilon=\frac{3\nu-2}{2\left(  3\nu-2\right)  -\kappa^{2}},\label{42}%
\end{equation}
and $r_{0}=r(\alpha/2)$ is the contact line distance.

For special opening angle $\pi$, it is straightforward to obtain the
streamline equation and the velocity components in terms of Cartesian
coordinates from Eqs.\ (\ref{301}), (\ref{34}), and (\ref{35}):
\begin{equation}
x=r_{0}\label{44}%
\end{equation}%
\begin{equation}
v_{x}=0\text{, \ \ \ }v_{y}=-\frac{1}{C\left(  \pi\right)  }\frac{J_{0}}{\rho
}.\label{45b}%
\end{equation}
Accordingly, we have $\kappa^{2}\left(  \pi\right)  =1$ and $\epsilon(\pi)=1$.
The streamline configuration and velocity are fully consistent with the
symmetry of the system, since the position of the apex is no longer well
defined when $\alpha=\pi$.

\begin{figure}
[ptb]
\begin{center}
\includegraphics[
trim=0.000000in 0.000000in -0.010408in 0.064025in,
natheight=3.423800in,
natwidth=4.163200in,
height=3.2214in,
width=3.5328in
]%
{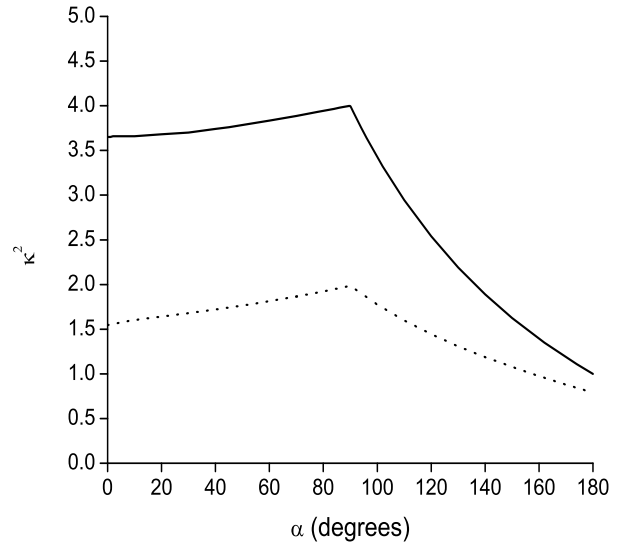}%
\caption{Dependence of parameter $\kappa^{2}$ on opening angle $\alpha$. The
solid line corresponds to the case of uniform evaporation. The dotted line,
obtained in Ref. 6, corresponds to the dry-surface evaporation with quadratic
profile (Eq. (25) in Ref. 6). }%
\label{kappa}%
\end{center}
\end{figure}

\begin{figure}
[ptbptb]
\begin{center}
\includegraphics[
trim=0.000000in 0.000000in 0.134809in 0.000000in,
natheight=3.338200in,
natwidth=4.122600in,
height=3.218in,
width=3.5328in
]%
{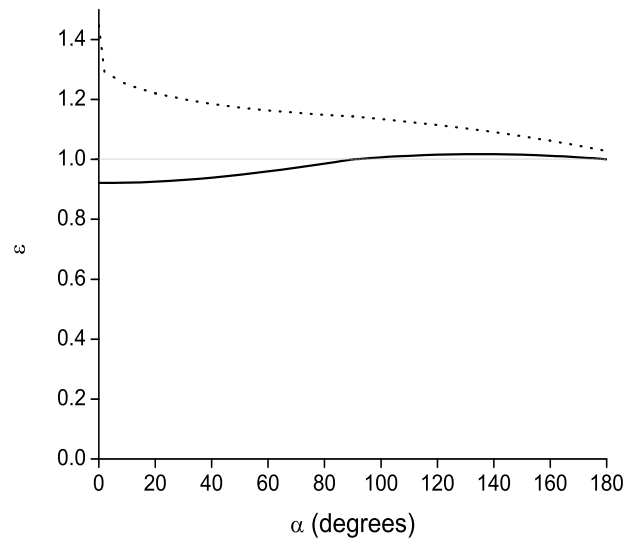}%
\caption{Dependence of the stream line asymptotic exponent $\epsilon$ (when
$\varphi\rightarrow0$) on the opening angle $\alpha$. The solid line
corresponds to the uniform evaporation case. The dotted line, obtained in Ref.
6, corresponds to the dry-surface evaporation with quadratic profile (Eq. (25)
in Ref. 6).}%
\label{epsilon1}%
\end{center}
\end{figure}

We do not have exact analytic results for exponents $\kappa^{2}\left(
\alpha\right)  $ and $\epsilon\left(  \alpha\right)  $ for arbitrary opening
angle $\alpha$, since equation (\ref{17}) can not be solved in closed form to
obtain $\tilde{\psi}(0)$. One has to solve the boundary value problem
(\ref{A13}) numerically to fix $\tilde{\psi}(0)$, and determine $\kappa^{2}$
and $\epsilon$ in terms of the opening angle $\alpha$. We show the dependence
of $\kappa^{2}$ and $\epsilon$ on the opening angle $\alpha$ in
Figs.\ \ref{kappa} and \ref{epsilon1}; for comparison, we also include results
found in Ref.\ 6 for the case of dry-surface evaporation. Numerical results
are in agreement with the analytic result we have found for the special case
$\alpha=\pi$.%

The exponent $\epsilon$ determines the asymptotic behavior of the streamlines
near the bisector $\varphi=0$. According to Eq.\ (\ref{41}), the distance
between a streamline and the bisector scales with $\varphi$ as $\varphi
r(\varphi)\propto\varphi^{1-\epsilon}$ in this limit. As is apparent in
Fig.\ \ref{epsilon1}, in the case of wet-surface evaporation $\epsilon$ is
equal to unity for $\alpha=\pi/2$ and $\pi$, and hence $\varphi r\left(
\varphi\right)  $ remains constant asymptotically. Geometrically, this means
that streamlines run parallel to the bisector when $\varphi\rightarrow0$. This
result also follows directly from our analytic solution (\ref{44}) for
$\alpha=\pi$ and from Fig.\ \ref{flowfield} for $\alpha=\pi/2$. For
$\alpha=\pi$ this geometric property of streamlines is well anticipated from
the symmetry of the system; for $\alpha=\pi/2$, however, it is not obvious.
For acute opening angles, we find $\epsilon<1$, and therefore the asymptotic
distance decreases when $\varphi\rightarrow0$, and the streamlines converge
toward the bisector as $r\rightarrow\infty$. The incoming particles are moving
along the trajectory away from the bisector. For obtuse opening angles, we
have $\epsilon>1$, and therefore the asymptotic distance increases when
$\varphi\rightarrow0$, and the streamlines diverge away from the bisector as
$r\rightarrow\infty$. Now the incoming particles first move along the
trajectory toward the bisector, reach a minimal distance, and then turn away
toward the contact line. For dry-surface evaporation \cite{6}, streamlines
always diverge away from the bisector, since $\epsilon>1$ in this case for all
opening angles (Fig.\ \ref{epsilon1}).

\subsection{Solute transfer and deposit growth}

As in the dry-surface evaporation case (Eqs.\ (55) and (56) in Ref.\ 6), one
can now calculate the time it takes for the solute particles initially located
at $\left(  r,\varphi\right)  $ (where $0\leq\varphi\leq\alpha/2$) to move
along the streamline to the contact line $(r_{0,}\alpha/2)$. We use
Eqs.\ (\ref{35}) and (\ref{37}) to write:
\begin{align}
t  & =\int_{\varphi}^{\alpha/2}\frac{rd\varphi}{v_{\varphi}}\label{47}\\
& =t_{0}\int_{\varphi}^{\alpha/2}\frac{\exp\left(  \nu\left(  3\nu-2\right)
\int_{\varsigma}^{\alpha/2}\tilde{\psi}\left(  \xi\right)  \left(
\frac{d\tilde{\psi}}{d\xi}\right)  ^{-1}d\xi\right)  }{\tilde{h}^{2}%
\frac{d\tilde{\psi}}{d\varsigma}}d\zeta,\nonumber
\end{align}
where
\begin{equation}
t_{0}=\frac{\rho}{J_{0}}\frac{r_{0}^{\nu}}{R(t)^{\nu-1}}\rightarrow\frac{\rho
}{J_{0}}\frac{r_{0}^{\nu}}{R_{i}^{\nu-1}}\label{48}%
\end{equation}
is a combination of the system parameters that has a dimension of time. In the
early stages of the drying process, we can approximate $R(t)$ by $R_{i}$, and
$t_{0}$ is therefore independent of time. For each streamline indexed by
contact line distance $r_{0}$, and for each time $t$, there exists a unique
$\varphi(r_{0},t)$ determined by Eq.\ (\ref{47}), such that all the solute
located in the area bounded by neighboring streamlines indexed by $r_{0}$ and
$r_{0}+dr_{0}$ in the domain $\varphi(r_{0},t)\leq\zeta\leq\alpha/2$ reaches
the contact line and becomes part of the deposit within time $t$. Let us
denote the mass accumulated between $r_{0}$ and $r_{0}+dr_{0}$ at the contact
line by time $t$ as $dm(r_{0},t)$.%

\begin{figure}
[ptb]
\begin{center}
\includegraphics[
height=3.1825in,
width=4.1001in
]%
{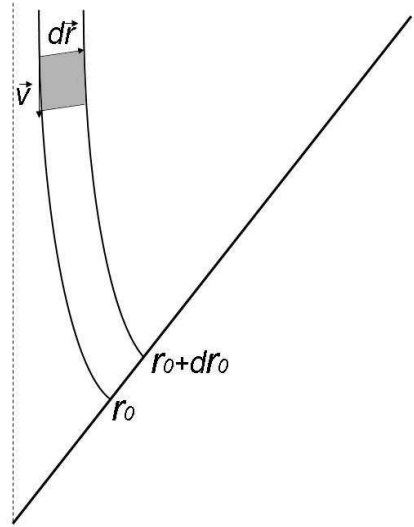}%
\caption{Qualitative sketch of the local approach. The solute in the shaded
area will be arriving at the contact line through the flowtube at time $t$,
and this corresponds to $d\left(  dm/dt\right)  (r_{0,}t)$. }%
\label{tube}%
\end{center}
\end{figure}

To study solute transfer and deposit growth, we want to understand the deposit
distribution along the contact line, as well as its growth rate with time. We
can consider the amount of solute $d(dm)/dt$ arriving at the contact line
during time $dt$ through the flow tube bounded by neighboring streamlines
indexed by $r_{0}$ and $r_{0}+dr_{0}$, as shown in Figure \ref{tube} (compared
to Ref.\ 6, where a global approach was employed, and the integral quantity
with respect to time $dm$ was computed). If we assume that the initial
concentration of the solute is constant $c_{0}$ everywhere in the drop, then
$d(dm)/dt$ can be expressed as:
\begin{equation}
\frac{d(dm)}{dt} (r_{0},t) = c_{0}h(\vec{r}(r_{0},\varphi(r_{0},t)))\left\vert
\vec{v}\mathbf{(}\vec{r})\times d\vec{r}\right\vert .\label{4801}%
\end{equation}

Again, we can start with the special opening angle $\pi$, for which analytical
results can be obtained straightforwardly. In Cartesian coordinates, the
surface shape (\ref{14b}) reads as $h=C(\pi)y$, the velocity is given by
Eq.\ (\ref{45b}), and $y=v_{y}t$. Expression (\ref{4801}) can be simplified
as
\begin{equation}
\frac{d(dm)}{dt} (r_{0},t)=c_{0}h(\vec{r}(r_{0},\varphi(r_{0},t)))\left\vert
v_{y}\right\vert dr_{0},\label{55a}%
\end{equation}
from where a power law can be found:
\begin{equation}
\frac{d}{dt}\left(  \frac{dm}{dr_{0}}\right)  =\frac{c_{0}}{C\left(
\pi\right)  }\left(  \frac{J_{0}}{\rho}\right)  ^{2}t\text{.}\label{57}%
\end{equation}
The deposition rate for $\alpha=\pi$ does not depend on $r_{0}$, which is
anticipated, since the position of the apex is no longer well defined. Thus,
for $\alpha=\pi$ there is a unique power law in the whole domain of $\varphi$
and for all times $t$ in the early drying stages.

For arbitrary opening angles $\alpha$ other than $\pi$, however, the
deposition rate can not be calculated analytically without knowledge of the
closed form of $\tilde{\psi}(\varphi)$. Instead, we have to analyze (\ref{47})
as well as other relevant quantities asymptotically in two different limiting
cases: $\varphi\rightarrow0$ and $\varphi\rightarrow\alpha/2$. These two
different asymptotic regions correspond to two different time regimes, when
the deposition rate follows different power laws, as was first introduced in Ref.\ 6:

\textbf{early time regime}: $t\ll t_{0}$, when only solute particles initially
located near the boundary can reach the contact line and become part of the
deposit. Properties of deposition in the time regime are governed by the
asymptotic $\varphi\rightarrow\alpha/2.$

\textbf{intermediate time regime}: $t_{0}\ll t\ll t_{f}$, when solute
particles initially located near the bisector are able to reach the contact
line. This time regime is governed by the limit $\varphi\rightarrow0$. The
condition $t\ll t_{f}$, where $t_{f}$ is the total drying time, means that we
are still considering early enough drying stages, where our model applies.

As argued in Ref.\ 6, the separation of two time regimes works worse when the
opening angle $\alpha$ increases toward $\pi$. This can be readily seen in our
case: as shown in (\ref{57}), the deposition rate follows the same power law
throughout the early drying stages when $\alpha=\pi$, and the two time regimes
are indistinguishable.

\subsubsection{Deposit growth in the early time regime}

In the early time regime, only the solute particles initially located near the
contact line contribute to the mass deposition.

As shown earlier, the velocity component $v_{r}$ vanishes in the limit
$\varphi\rightarrow\alpha/2$, and streamlines are perpendicular to the contact
line. Expression (\ref{4801}) has a very simple form:
\begin{equation}
\frac{d(dm)}{dt}\propto v_{\varphi}hdr_{0},\qquad\varphi\rightarrow
\frac{\alpha}{2}.\label{6501}%
\end{equation}
In this limit, the asymptotic forms are $h\propto r_{0}^{\nu}(\alpha
/2-\varphi)$ (Eq. (\ref{13})), $v_{\varphi}\propto r_{0}^{1-\nu}$ (Eq.
(\ref{35})), and $t(\varphi)\propto r_{0}^{\nu}(\alpha/2-\varphi)$ (Eq.
(\ref{47})), and therefore Eq.\ (\ref{6501}) can be written in the form of a
power law:
\begin{equation}
\frac{d}{dt}\left(  \frac{dm}{dr_{0}}\right)  \propto tr_{0}^{\beta
},\label{6502}%
\end{equation}
where
\begin{equation}
\beta=1-\nu.\label{66}%
\end{equation}

\begin{figure}
[ptb]
\begin{center}
\includegraphics[
trim=0.000000in 0.000000in 0.008538in -0.019516in,
natheight=3.423800in,
natwidth=4.065500in,
height=3.1012in,
width=3.5034in
]%
{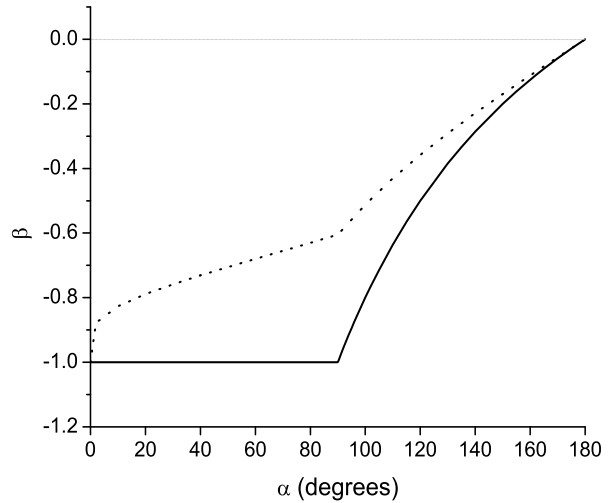}%
\caption{Dependence of the exponent $\beta$ of the contact line distance
$r_{0}$ on the opening angle $\alpha$ in the power law $dm/dr_{0}$
(Eq.\ (\ref{65})) in the early time regime. The solid line corresponds to the
uniform evaporation, and the dotted line corresponds to the dry-surface
evaporation with the quadratic profile (Eq.\ (25) in Ref.\ 6).}%
\label{betta}%
\end{center}
\end{figure}

One can also use the approach employed in Ref.\ 6 to find the power law of the
deposition rate (compared to Eq.\ (60) in Ref.\ 6):
\begin{equation}
\frac{dm}{dr_{0}}(r_{0,}t)\propto t^{2}r_{0}^{\beta},\label{65}%
\end{equation}
where $\beta$ is again given by (\ref{66}), which is in agreement with result
(\ref{6502}) and the result we found for special angle $\pi$ (\ref{57}). The
dependence of the exponent $\beta$ on the opening angle is shown in
Fig.\ \ref{betta}, where the exponent obtained in Ref.\ 6 for the case of
dry-surface evaporation is also included for comparison.

The relation (\ref{66}) can be understood in the following way. Let $\delta$
be the distance from the contact line. When $\delta$ is small, mass
conservation demands $h\mathbf{v} \propto J\delta$, and hence $\mathbf{v}
\propto(dh/d\delta)^{-1}$, i.e., near the contact line the velocity should be
inversely proportional to the slope of the drop surface $\left\vert
dh/d\delta\right\vert $. According to (\ref{13}), near the contact line $h =
r^{\nu} \tilde{h}\left( \varphi\right)  \approx r^{\nu}\tilde{h}%
(\alpha/2-\delta/r)$, and $\left\vert dh/d\delta\right\vert \propto r^{\nu-1}
\left.  (d\tilde{h}/d\varphi) \right\vert _{\varphi= \alpha/2}$. Therefore,
the velocity is proportional to $r^{1-\nu}$, so is the mass deposition rate,
and Eq.\ (\ref{66}) follows.

\subsubsection{Deposit growth in the intermediate time regime}

In the intermediate time regime, we need again to find the power law of
$d(dm)/dt$. By analyzing expressions (\ref{13}), (\ref{34}), (\ref{35}),
(\ref{41}), and (\ref{47}) in the limit $\varphi\rightarrow0$, we find that
relevant physical quantities assume the following asymptotic forms:
\begin{equation}
r\propto r_{0}\varphi^{-\epsilon},\qquad h\propto r_{0}^{\nu}\varphi
^{-\nu\epsilon},\label{7001}%
\end{equation}%
\begin{equation}
v_{r}\propto-r_{0}^{1-\nu}\varphi^{-\epsilon(1-\nu)},\qquad v_{\varphi}\propto
r_{0}^{1-\nu}\varphi^{1-\epsilon(1-\nu)},\label{7004}%
\end{equation}%
\begin{equation}
t\propto r_{0}^{\nu}\varphi^{-\nu\epsilon}.\label{7005}%
\end{equation}
Then, according to Eq. (\ref{4801}),
\begin{equation}
\frac{d(dm)}{dt}\propto h\left\vert \vec{v}(\vec{r})\times d\vec{r}\right\vert
=h\left\vert v_{\varphi}dr-v_{r}rd\varphi\right\vert .\label{7006}%
\end{equation}
By assuming $dt=0$ along the direction of $d\vec{r}$, we can derive the
relation $d\varphi\propto-\left(  \varphi/r_{0}\right)  dr_{0}$ from
Eq.\ (\ref{7005}). Using Eqs.\ (\ref{7001}) and (\ref{7004}), expression
(\ref{7006}) can be simplified as:
\begin{equation}
\frac{d}{dt}\left(  \frac{dm}{dr_{0}}\right)  \propto r_{0}\varphi
^{1-2\epsilon}.\label{7007}%
\end{equation}
Together with (\ref{7005}), we finally obtain:%
\begin{equation}
\frac{d}{dt}\left(  \frac{dm}{dr_{0}}\right)  \propto t^{\delta-1}%
r_{0}^{\gamma},\label{7008}%
\end{equation}
where
\begin{equation}
\delta=1+\frac{\kappa^{2}}{\nu\left(  3\nu-2\right)  }\label{71}%
\end{equation}
and
\begin{equation}
\gamma=1-\frac{\kappa^{2}}{3\nu-2}.\label{73}%
\end{equation}

\begin{figure}
[ptb]
\begin{center}
\includegraphics[
trim=0.000000in 0.000000in -0.105538in -0.138322in,
natheight=3.423800in,
natwidth=4.122600in,
height=3.2621in,
width=3.563in
]%
{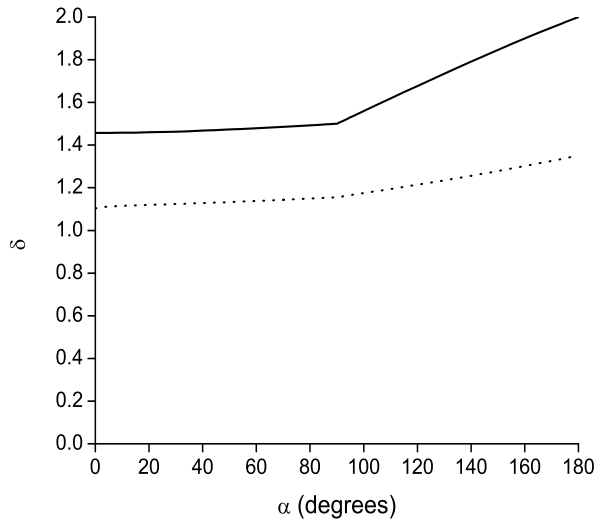}%
\caption{Dependence of the exponent $\delta$ of time $t$ on the opening angle
$\alpha$ in the power law $dm/dr_{0}$ (Eq.\ (\ref{70})) in the intermediate
time regime. The solid line corresponds to the uniform evaporation, and the
dotted line corresponds to the dry-surface evaporation with the quadratic
profile (Eq.\ (25) in Ref.\ 6).}%
\label{delta}%
\end{center}
\end{figure}

\begin{figure}
[ptbptb]
\begin{center}
\includegraphics[
trim=0.000000in 0.000000in 0.103801in 0.020201in,
natheight=3.423800in,
natwidth=4.135500in,
height=3.2214in,
width=3.5328in
]%
{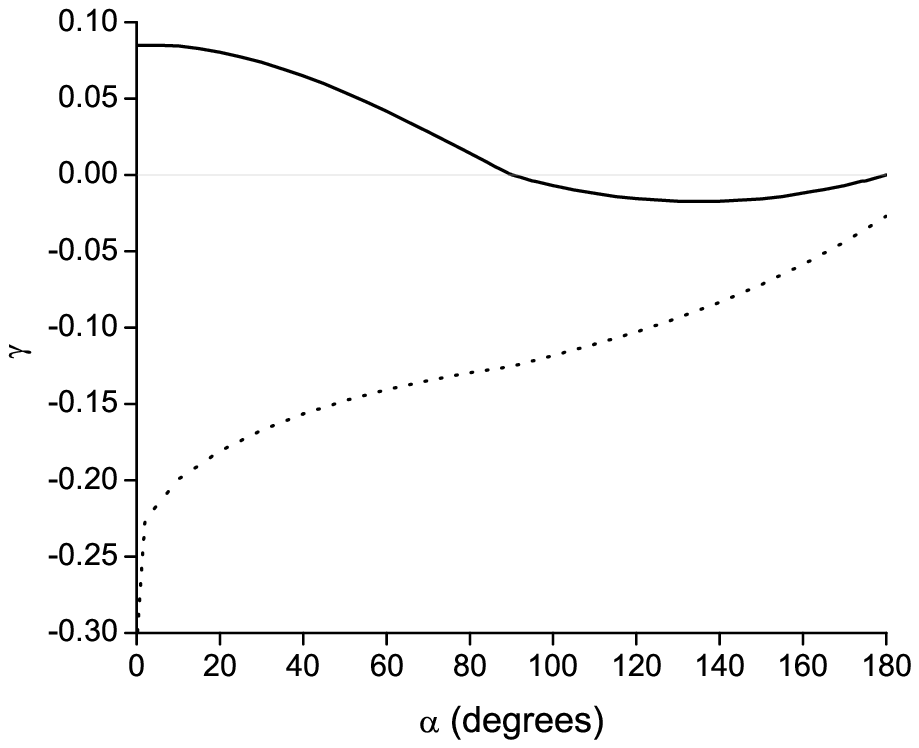}%
\caption{Dependence of the exponent $\gamma$ of the contact line distance
$r_{0}$ on the opening angle $\alpha$ in the power law $dm/dr_{0}$
(Eq.\ (\ref{70})) in the intermediate time regime. The solid line corresponds
to the uniform evaporation, and the dotted line corresponds to the dry-surface
evaporation with the quadratic profile (Eq.\ (25) in Ref.\ 6).}%
\label{gamma}%
\end{center}
\end{figure}

In this limit, we can also follow Ref.\ 6 to compute the power law of
$dm/dr_{0}$, and we find the deposition rate to be (compared to Eq.\ (63) in
Ref.\ 6):%
\begin{equation}
\frac{dm}{dr_{0}}\left(  r_{0},t\right)  \propto t^{\delta}r_{0}^{\gamma
},\label{70}%
\end{equation}
with the same scaling exponents given by (\ref{71}) and (\ref{73}). Again
Eqs.\ (\ref{7008}) and (\ref{70}) are in agreement with each other and
exponents (\ref{71}) and (\ref{73}) are in agreement with the values of
exponents we obtained for special angle $\pi$. We show the dependence of the
exponents $\delta$ and $\gamma$ on the opening angle in Figs.\ \ref{delta} and
\ref{gamma}, where we also include results of Ref.\ 6 for comparison.%

\section{Comparison with the dry-surface evaporation case}

\begin{figure}
[ptb]
\begin{center}
\includegraphics[
natheight=3.338200in,
natwidth=4.122600in,
height=3.0511in,
width=3.5163in
]%
{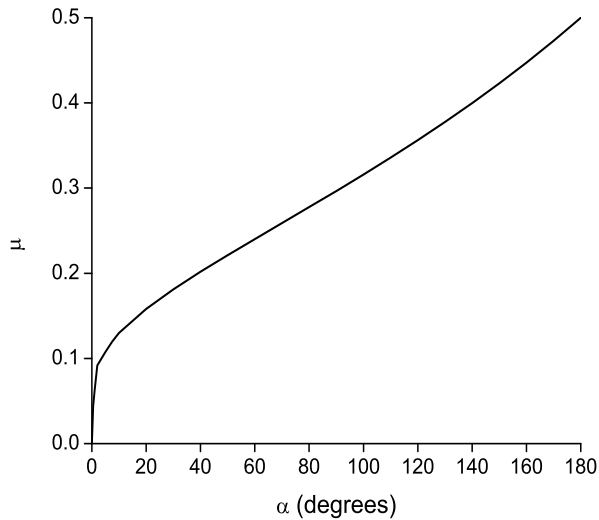}%
\caption{Relation between the exponent $\mu$ and the opening angle $\alpha$,
in the case of diffusion-controlled evaporation from an angular region}%
\label{newmu}%
\end{center}
\end{figure}

Popov and Witten \cite{6} considered a general evaporation rate $J(r,\varphi)
$ of the asymptotic form
\begin{equation}
J \rightarrow r^{\mu-1}\left( \frac{\alpha}{2}-\left\vert \varphi\right\vert
\right) ^{-\lambda}, \qquad\left\vert \varphi\right\vert \rightarrow
\frac{\alpha}{2}.\label{-3}%
\end{equation}
The wet-surface case considered here corresponds to $\mu=1$ and $\lambda=0$.
The results of the previous section may be obtained by setting $\mu=1$ and
$\lambda=0$ in the general expressions of Ref.\ 6. In some cases this limit
permits simpler expressions and more explicit solutions as we have seen. For
the dry-surface evaporation case studied in Ref.\ 6, $\lambda=1/2$ and $\mu$
is an explicit function of the opening angle $\alpha$ (Fig.\ \ref{newmu})
reflecting the singular behavior of the Laplacian vapor concentration field
\cite{9}. Compared to the dry-surface evaporation, the uniform evaporation
rate yields different deposition properties in both early and intermediate
time regimes.%

\subsection{Early time regime}

In this regime, only the asymptotic form (\ref{-3}) of the evaporation rate
$J$ when $\varphi$ approaches $\pm\alpha/2$ matters. While $\mu$ is relevant
to the dependence of the reduced pressure function $\psi$ on coordinate $r$,
the exponent $\lambda$ governs the singular property of the reduced pressure
function near the contact line (a direct observation is that, due to nonzero
value of $\lambda=1/2$ in the dry-surface evaporation, the order of divergence
of $\tilde{\psi}$ is higher by $1/2$ in that case). For the deposition rate in
the early time regime, the power law found in the dry-surface evaporation case
\cite{6}, as well as our result (\ref{65}), is uniquely determined by the
local singular properties of the reduced pressure function at the contact
line, and those power law exponents have simple algebraic expressions in terms
of $\mu$, $\lambda$, and $\nu$.

Both results of Ref.\ 6 and our results show that the exponent of time $t$ in
the power law of deposition rate is independent of the opening angle $\alpha$
in the early time regime, although in our case the deposition process goes
slower with exponent $2$ instead of $4/3$ in Ref.\ 6. As regards the
dependence on the contact line distance $r_{0}$, both results of Ref.\ 6 and
our results (Fig.\ \ref{betta}) show that $\beta$ always remains between $-1$
and $0$ for obtuse opening angles $\alpha$, and the integrability of the
singularity at $r_{0}=0$ holds. As argued in Ref.\ 6, despite having larger
deposition rate, the vertex of the sector does not dominate the sides, and the
deposition accumulation at the vertex is not qualitatively different from the
deposition accumulation on the sides for obtuse opening angles.

For acute opening angles $\alpha$, although $\beta$ remains between $-1$ and
$0$ for the dry-surface evaporation, it is constantly $-1$ in our case. It
seems that we have a more concentrated deposition pattern with a nonsingular
form of the evaporation rate. The singular flow of the dry-surface evaporation
rate $J$ may deflect the streamlines towards the contact line more than in our
case, and hence drive more fluid and mass to the sides and thus away from the
apex.\footnote{We actually plotted the streamlines for both dry-surface and
wet-surface evaporation cases, however, we did not notice any substantial
qualitative difference. This result is important and should be provable
experimentally, since one should be able to measure mass accumulation at the
sides and at the vertex and compare the two evaporation cases.}

Furthermore, the $1/r_{0}$ dependence of the deposition rate in the early time
regime for acute opening angles seems to violate the integrability and
suggests a logarithmic divergence: an arbitrarily large fraction of the mass
may accumulate within an arbitrarily small distance from the apex. To resolve
this possible singularity, we note that $1/r_{0}$ dependence occurs not at all
times, but only at early times, which are almost never observed near the
vertex. Indeed, for each time $t$ one can define a crossover length
\begin{equation}
r^{\ast}(t)=\left(  tR_{i}^{\nu-1}\frac{J_{0}}{\rho}\right)  ^{1/\nu
}.\label{-102}%
\end{equation}

According to the definition of the time scale $t_{0}$ (\ref{48}), the early
time condition $t\ll t_{0}$ corresponds to the condition $r_{0}\gg r^{\ast
}(t)$ in terms of the location of that regime along the contact line. Thus, at
each time $t$, the early time regime is observed only away from the vertex,
and the areas near the vertex are always in the intermediate time regime. The
natural cutoff $r^{\ast}(t)$ actually saves the mass deposition at the tip
from being logarithmically infinite.

\subsection{Intermediate time regime}

Distinctive power laws characterizing this regime (such as the configuration
of the streamlines and the deposition rate) depend on the entire range of
$\varphi$, including the limit $\varphi\rightarrow0$, which dominates
properties of these power laws. An important parameter in this regime is the
value of the reduced pressure function at the bisector $\tilde{\psi}(0)$, or
the parameter $\kappa^{2}$ (Eq.\ (\ref{40})). Near the bisector, both the
evaporation rate $J$ and function $\tilde{\psi}$ are regular, and in order to
find $\tilde{\psi}(0)$ one needs to solve the differential equation (\ref{17})
in the full domain of $\varphi$ from $0$ to $\alpha/2$. Therefore, in the
intermediate time regime relevant quantities depend on the functional form of
the evaporation rate $J$ and surface shape $h$ in the entire domain of
$\varphi$ and require solution of the main equation (\ref{17}), in contrast to
the early time regime where only the singular behavior at the contact line matters.

The most interesting result in this regime is certainly the qualitatively
different behavior of physical quantities in the geometry of the acute opening
angles versus obtuse opening angles. Our numerical result for $\kappa^{2}$
(Fig.\ \ref{kappa}) is not qualitatively different from that found in Ref.\ 6,
as $\kappa^{2}$ changes dramatically at $\alpha=\pi/2$ in both cases. This is
understandable, because the non-smoothness at $\alpha=\pi/2$ is uniquely due
to the crossover of the exponent $\nu$ in the leading order term determining
the surface shape (Eqs.\ (\ref{14a}) and (\ref{14b})), while the evaporation
rate $J$ behaves smoothly in both cases. However, changes for other quantities
that take place at the right opening angle are qualitative in the wet-surface
case, unlike in the dry-surface case.

The flow field configuration near the bisector as $\varphi\rightarrow0$, is
governed by the exponent $\epsilon$ shown in Fig.\ \ref{epsilon1}. In the case
of dry-surface evaporation, $\epsilon$ is larger than $1$, except for
$\alpha=\pi$, and the streamlines asymptotically diverge from the bisector. In
contrast, our wet-surface results show that the streamlines asymptotically
converge to the bisector ($\epsilon<1$) for acute opening angles, while for
obtuse opening angles they still diverge. In particular, for the critical
angle $\pi/2$ streamlines run parallel to the bisector line, which is not
anticipated intuitively.

For the power law of the deposition rate in the intermediate time regime, we
obtained the asymptotic form (\ref{70}). The result for the exponent
$\delta\left( \alpha\right) $ (Fig.\ \ref{delta}) follows the same qualitative
pattern as the result in Ref.\ 6, although it is everywhere larger, so that
the deposition process goes slower, like in the early time regime.

For dry-surface evaporation, $\gamma(\alpha)$ remains negative, monotonically
increases with $\alpha$, and becomes $0$ at $\alpha=\pi$ (compatible with the
symmetry). In the wet-surface evaporation case, $\gamma(\alpha)$ follows a
richer pattern. Physically, as $\gamma$ is strictly larger than $-1$, the
integrability property holds. However, while $\gamma$ remains negative for
obtuse opening angles, it becomes positive for acute angles, and therefore the
deposition accumulation is in favor of the sides rather than the vertex as in
Ref.\ 6. In particular, as $\gamma$ goes to $0$ at $\alpha=\pi/2$ and its
absolute value remains small nearby, it seems that the wet-surface evaporation
case yields a relatively uniform deposition pattern for a wide range of
opening angles in the intermediate time regime.

The exponent $\gamma$ controlling mass deposition is closely related to the
trajectory exponent $\epsilon$. From expressions (\ref{42}) and (\ref{73}) one
can obtain a simple relation:%
\begin{equation}
\epsilon=\frac{1}{1+\gamma},\label{80}%
\end{equation}
which is also implicit in Ref.\ 6. Intuitively, in the intermediate time
regime, virtually all the solute between the bisector and the contact line is
swept to the contact line and becomes part of the deposit, and therefore
$\gamma$, the exponent of the contact line distance $r_{0}$ indexing the
streamlines, should be related to the geometric distribution of the
streamlines near the bisector away from the vertex. For $\alpha=\pi$ and
$\pi/2$, exponent $\epsilon=1$, and the streamlines are uniformly distributed
near the bisector. Therefore, the solute is uniformly carried to the contact
line by the flow, and the deposition rate should be independent of the contact
line distance, hence $\gamma=0$. For other angles, the streamlines become
unevenly distributed away from the vertex. When $\epsilon>1$, they diverge
away from the bisector as $\varphi\rightarrow0$, the smaller contact line
distance $r_{0}$ is, the nearer the corresponding streamline to the bisector,
and more solute will be carried to the spot along the streamline. Therefore
the deposition is in favor of the vertex, and $\gamma$ is negative. When
$\epsilon<1$, $\gamma$ is positive by the same argument.

To be more precise mathematically, in the intermediate time regime $t_{0}\ll
t$ the deposition is uniquely determined by the streamline configuration near
the bisector line (this is in contrast to the early time regime, where the
deposition rate is closely related to the slope of the surface shape near the
contact line). The amount of solute deposited near $r_{0}$ is controlled by
the width of the gap between adjacent streamlines indexed by $r_{0}$ and
$r_{0}+\Delta r_{0}$ near the bisector line, which is proportional to
$r\Delta\varphi$ in the limit $\varphi\rightarrow0$. Near the bisector line,
the streamline indexed by $r_{0}$ can be expressed as $r\propto r_{0}%
\varphi^{-\epsilon}$ (Eq.\ (\ref{7001})). If we consider a small patch
$r=const$ near the bisector, and study the intersections of \ those
streamlines with this patch, we have
\begin{equation}
0=dr\propto\varphi^{-\epsilon}dr_{0}-\epsilon r_{0}\varphi^{-\epsilon
-1}d\varphi,\label{80b}%
\end{equation}
and therefore
\begin{equation}
\frac{d\varphi}{dr_{0}}=\frac{\varphi}{\epsilon r_{0}}.\label{80c}%
\end{equation}
According to the above argument, Eqs.\ (\ref{7001}) and (\ref{80c}) yield
\begin{equation}
\frac{dm}{dr_{0}}\propto\frac{d\varphi}{dr_{0}}\propto\frac{1}{\epsilon}%
r_{0}{}^{1/\epsilon-1},\label{80d}%
\end{equation}
and the relation (\ref{80}) follows immediately by comparison of the last
expression with Eq.\ (\ref{70}).

\section{Discussion}

The theoretical framework, first established in Ref.\ 6 and studied here in
continuation, captures the essential mechanism of the deposit growth, but does
not take into account a number of additional effects that can modify the
deposition, and some restrictions and shortcomings remain \cite{6}.

The crossover at the opening angle $\alpha=\pi/2$ (as manifested in the
surface shape $h$ (\ref{13}), (\ref{14a}) and (\ref{14b})), which is certainly
independent of the evaporation rate, is quite a subtle point of the theory. As
shown in Appendix B, in principle, modifications to the results obtained in
this work as well as in Ref.\ 6 by the asymptotic analysis are needed in the
neighborhood of the right opening angle $[\pi/2-\Delta\alpha,\pi
/2+\Delta\alpha]$, with $\Delta\alpha\sim(\pi/4)\left\vert \ln(r/R_{i}%
)\right\vert ^{-1}$ (Eq.\ (\ref{116})). The thinner and flatter the drop is,
and the closer to the apex, the better our results apply.

Further, we believe the power law exponents obtained by the asymptotic
analysis are exact, except for a possible logarithmic modification at
$\alpha=\pi/2$, where crossover of exponents could happen. Exact properties of
the other experimentally testable physical quantities (the velocity
components, the reduced pressure function $\tilde{\psi}$) in the neighborhood
$[\pi/2-\Delta\alpha,\pi/2+\Delta\alpha]$ could be in principle interpolated,
as shown in Appendix B, since all the physical properties of the system should
depend on the opening angle $\alpha$ continuously.

Another interesting observation is the form of the evaporation rate. In this
paper, we considered the uniform evaporation, which is simpler than the
dry-surface evaporation studied in Ref.\ 6. The form of the evaporation rate
is not arbitrary, and apart from other physical restrictions, it should be
compatible with the symmetry of the system. One such consideration would be
the following: in the limit $\alpha\rightarrow\pi$ the position of the apex is
no longer well defined, and the deposit rate should not depend on the contact
line distance $r_{0}$ in both early and intermediate time regimes; therefore
$\beta(\pi)=\gamma(\pi)=0$. Combined with the general expression for $\beta$
(Eq.\ (61) in Ref.\ 6), this condition demands
\begin{equation}
\lambda(\pi)+\mu(\pi)=1,\label{1000}%
\end{equation}
which is satisfied in both cases:\footnote{The physical origin of this
condition is very simple: for the opening angle $\alpha=\pi$, the evaporation
rate should depend only on the normal distance to the contact line
$r\sin(\alpha/2-\left\vert \varphi\right\vert )$, which reduces to
$r(\alpha/2-\left\vert \varphi\right\vert )$ when $\left\vert \varphi
\right\vert \rightarrow\alpha/2$. Thus, the exponents of $r$ and
$(\alpha/2-\left\vert \varphi\right\vert )$ must be equal for this opening
angle, and the general expression (\ref{-3}) immediately yields $\mu
-1=-\lambda$.} for the dry-surface evaporation $\lambda\equiv1/2$ and $\mu
(\pi)=1/2$, and for the uniform evaporation $\lambda\equiv0$ and $\mu\equiv1$.
Although $\gamma\left(  \pi\right)  =0$ in our case, this did not hold exactly
for the exponent $\gamma$ in the dry-surface case, since the approximation of
$J$ used for the numerical solution in Ref.\ 6 was not in exact conformity
with the symmetry. In contrast to $\beta$, where $\beta(\pi)=0$ can be traced
back to the simple relation (\ref{1000}), the exponent $\gamma$ is related to
the parameter $\kappa^{2}$, which depends on the solution of the main equation
(\ref{17}) in the entire domain of $\varphi$. It is interesting to see how the
differential equation (\ref{17}), which is merely a statement of local
conservation of mass, together with the proper form of the evaporation rate,
yields the deposition properties in full conformity with the symmetry of the system.

In general, various properties of the evaporating drop in the intermediate
time regime depend on the mathematical structure over the whole domain of
$\varphi$, and more thorough analytical treatment of the main equation
(\ref{17}) is certainly appealing. The criticality of the opening angle
$\alpha=\pi/2$ demands extra attention: Why do some physical properties differ
between acute angle and obtuse angle? Why was not this separation so apparent
in Ref.\ 6, where the same $h$ entered? How is this separation related to the
form of the evaporation rate? Naively, these questions can be readily
addressed by saying that the criticality of the right angle is uniquely due to
the crossover of the leading terms in the full expansion of the surface shape
$h$ at $\alpha=\pi/2$. When the evaporation rate is uniform, it does not
introduce any further singularity, and thus helps to retain the trace of the
criticality of $h$ in the resulting physical quantities and phenomena. In the
case of the dry-surface evaporation, the stronger dependence of $J$ on
coordinates may overshadow the coordinate dependence of $h$, and the latter
may appear less significant in the results. In particular, it seems that the
change of the exponent $\mu$ in the evaporation rate $J$ is in such a
direction as to compensate for the change of exponent $\nu$ in the surface
height $h$. More mathematically rigorous treatment needs to be done to make
this argument clearer.

Experimentally, it is interesting to note some possible applications of our
results. Our work shows that wet-surface evaporation at early times with acute
opening angle achieves the greatest concentration towards the apex.
Accordingly, one could actually achieve a great concentration of mass by
allowing the evaporation to occur for a short time, then allowing the
dissolved solute to diffuse and equilibrate, then allowing another bit of
evaporation, and so forth. In this way, one could approach the behavior of
having a finite fraction of the mass within some small distance of the apex.
Another interesting aspect is the nearly perfect uniformity of the deposition
for the intermediate time regime with opening angle $\alpha\simeq\pi/2$. This
kind of uniform deposition may be useful, especially when a small amount of
the concentrated substance is sufficient. For example, a dilute solution of
reagents can be concentrated strongly at the contact line, thereby inducing a
chemical reaction there. The evaporation mechanism assures that the
concentration is a known function of the position and the initial dilution.
Likewise, trace amounts of solute can be rendered more easily detectable by
causing them to concentrate at a contact line.

\section{Conclusion}

The wet-surface evaporation of an angular drop yields surprisingly rich and
potentially useful behavior. This behavior complements the previously studied
work on dry-surface evaporation \cite{6}. Though our case lacks the
distinctive singular evaporation of the dry-surface evaporation case,
remarkably, it leads to a \textit{stronger} focusing of solute towards the
apex. Further, it can create two qualitatively different types of flow,
according to whether the opening angle is acute or obtuse. The deposition
profile is remarkably uniform for the intermediate times when the opening
angle is close to a right angle. Now that these deposition properties have
been established, they may well prove useful. For example, they provide a
means of concentrating trace solutes in a liquid in a rapid and quantitatively
predictable way. They also create distinctive capillary flow fields and
distinctive concentration profiles of solute. We expect this kind of
microscopic, singular, evaporative flow to play an increasing role in the
technology of small scale material synthesis, processing and analysis.

\begin{acknowledgments}
This work was supported in part by the National Science Foundation's MRSEC
Program under Award Number DMR-0213745.
\end{acknowledgments}

\appendix

\section{Reduced pressure equation: outer boundary condition and
regularization}

In this appendix, we provide a justification of the outer boundary condition
(\ref{222}), and we also show how the boundary problem can be regularized. In
Ref.\ 6, Eq.\ (\ref{17}) was solved using the global mass conservation
condition
\begin{equation}
\rho\int_{-\alpha/2}^{\alpha/2}\left\vert v_{r}\right\vert hrd\varphi=\int
_{0}^{r}\int_{-\alpha/2}^{\alpha/2}J_{0}rdrd\varphi,\label{20}%
\end{equation}
where $v_{r}=h^{2}\partial_{r}\psi$ is the radial component of the flow
velocity \cite{6}. However, Eq.\ (\ref{17}) in itself is an expression of
local mass conservation following Eq.\ (\ref{3}). It is therefore interesting
to find out explicitly how Eq.\ (\ref{20}) represents new information that can
restrict the solution. Here we explain the origin of the new information and
show that it takes the form of an explicit boundary condition.

With the expressions of $h$ and $\psi$, Eq.\ (\ref{20}) can be simplified as
\begin{equation}
\int_{0}^{\alpha/2}\left(  2\left(  3\nu-2\right)  \tilde{h}^{3}\tilde{\psi
}-1\right)  d\varphi=0.\label{21a}%
\end{equation}
The local mass conservation implicitly stated in Eq.\ (\ref{17}) allows us to
express $\tilde{\psi}$ in terms of the derivative $d\tilde{\psi}/d\varphi$:
\begin{equation}
2(3\nu-2)\tilde{\psi}\tilde{h}^{3}=\tilde{h}^{3}\frac{d^{2}\tilde{\psi}%
}{d\varphi^{2}}+3\tilde{h}^{2}\frac{d\tilde{h}}{d\varphi}\frac{d\tilde{\psi}%
}{d\varphi}+1.
\end{equation}
Then integration by parts, together with the boundary condition (\ref{19}),
converts the integral condition (\ref{21a}) into the boundary condition
(\ref{222}).

We now show that condition (\ref{222}) does in fact restrict the general form
of $\tilde{\psi}$ and fixes the order of its divergence near the contact line.
To demonstrate this, we study the asymptotic behavior of $\tilde{\psi}$. As
both $\tilde{h}$ and $\tilde{\psi}$ are even in $\varphi$, we can consider
only the domain $0\leq\varphi\leq\alpha/2$. We restrict our attention to the
limit $\varphi\rightarrow\alpha/2$ and keep in mind that $\tilde{h}%
\propto(\alpha/2-\varphi)$ in this limit. From the expressions of $\tilde{h}$
(\ref{14a}) and (\ref{14b}), as well as Eq.\ (\ref{17}), we see that
$\tilde{\psi}$ necessarily diverges in this limit, and the third term on the
left side of Eq.\ (\ref{17}), which is of the lowest order of divergence, can
be neglected in the asymptotic analysis. Thus, Eq.\ (\ref{17}) reduces to:
\begin{equation}
\frac{d^{2}\tilde{\psi}}{d\varphi^{2}}+\frac{3}{\tilde{h}}\frac{d\tilde{h}%
}{d\varphi}\frac{d\tilde{\psi}}{d\varphi}+\frac{1}{\tilde{h}^{3}}=0.\label{23}%
\end{equation}
Eq.\ (\ref{23}) can be solved analytically. First, consider the homogeneous
first order differential equation for $d\tilde{\psi}/d\varphi:$%
\begin{equation}
\frac{d}{d\varphi}\left(  \frac{d\tilde{\psi}}{d\varphi}\right)  +\frac
{3}{\tilde{h}}\frac{d\tilde{h}}{d\varphi}\frac{d\tilde{\psi}}{d\varphi
}=0,\label{23b}%
\end{equation}
which has the general solution of the form:
\begin{equation}
\frac{d\tilde{\psi}}{d\varphi}=c\frac{1}{\tilde{h}^{3}},\label{23d}%
\end{equation}
with $c$ being an arbitrary constant. In order to obtain the general solution
of the inhomogeneous equation (\ref{23}), we let $c$ be a function of
$\varphi$, i.e., $d\tilde{\psi}/d\varphi=c(\varphi)/\tilde{h}^{3}$, then plug
it into Eq.\ (\ref{23}) and find
\begin{equation}
\frac{dc}{d\varphi}=-1,\label{23c}%
\end{equation}
which means
\begin{equation}
c(\varphi)=-\varphi+ const.\label{23e}%
\end{equation}
Combining Eqs.\ (\ref{23d}) and (\ref{23e}), we find that Eq.\ (\ref{23}) has
a general solution of the form:
\begin{equation}
\tilde{\psi} \rightarrow C_{1} +C_{2} \int_{0}^{\varphi} \frac{d\xi}{\tilde
{h}^{3}(\xi)} - \int_{0}^{\varphi} \frac{\xi}{\tilde{h}^{3}(\xi)}d\xi,
\qquad\varphi\rightarrow\frac{\alpha}{2},\label{24}%
\end{equation}
where $C_{1}$ and $C_{2}$ are arbitrary constants. We can display the
divergence of $\tilde{\psi}$ in this limit more explicitly by expanding the
right-hand side of Eq.\ (\ref{24}) in terms of $(\alpha/2-\varphi)$:
\begin{align}
\tilde{\psi}  &  \rightarrow \frac{1}{2} \left( C_{2} - \frac{\alpha}%
{2}\right)  \left\vert \frac{d\tilde{h}}{d\varphi}\left( \frac{\alpha}%
{2}\right)  \right\vert ^{-3} \left( \frac{\alpha}{2} - \varphi\right)
^{-2}\nonumber\\
&  +  \left\vert \frac{d\tilde{h}}{d\varphi}\left( \frac{\alpha}{2}\right)
\right\vert ^{-3}\left( \frac{\alpha}{2} - \varphi\right) ^{-1} +
const,\label{101a}%
\end{align}
where we only retained the divergent terms and the constant term, and
accordingly,
\begin{align}
\frac{d\tilde{\psi}}{d\varphi}  &  \rightarrow \left( C_{2} - \frac{\alpha}%
{2}\right)  \left\vert \frac{d\tilde{h}}{d\varphi} \left( \frac{\alpha}%
{2}\right)  \right\vert ^{-3} \left( \frac{\alpha}{2} - \varphi\right)
^{-3}\nonumber\\
&  +  \left\vert \frac{d\tilde{h}}{d\varphi} \left( \frac{\alpha}{2}\right)
\right\vert ^{-3} \left( \frac{\alpha}{2} - \varphi\right) ^{-2}.\label{101b}%
\end{align}
Now it becomes immediately apparent that condition (\ref{222}) demands $C_{2}
= \alpha/2$, and only the first order divergence of $\tilde{\psi}$ is allowed
(which is always present, as is clear from Eq.\ (\ref{101a})).

On reflection, the physical content associated with the boundary condition
(\ref{222}) may invite further exposition. Taking into account Eqs.\ (\ref{9})
and (\ref{10}), we note that the boundary condition (\ref{222}) states
physically $h\mathbf{v}=0$ at the contact line. Consider a region within
distance $\delta$ from the boundary. The influx $h\mathbf{v}$ should be
balanced with the evaporation flux, which is proportional to $J\delta$. Let
$\delta$ go to $0$, and our result follows. Mathematically, one can argue that
singular behavior of Eq.\ (\ref{17}) and the order of divergence of
$\tilde{\psi}$ in the limit $\varphi\rightarrow\alpha/ 2$ are uniquely
determined by the term $1/\tilde{h}^{3}$. The solution with a higher-order
divergence, though compatible with the mathematical structure, is not allowed
by the physics.

To make the boundary condition (\ref{222}) easier to handle mathematically, we
define a regularized function:
\begin{equation}
\tilde{\chi}=\tilde{h}^{2}\tilde{\psi},\label{A121}%
\end{equation}
where $\tilde{h}^{2}$ is introduced to compensate for the second order
divergence in $\tilde{\psi}$ at the contact line allowed by Eq.\ (\ref{17}).
The original problem is converted to the standard boundary value problem:
\begin{equation}
\tilde{h}\frac{d^{2}\tilde{\chi}}{d\varphi^{2}}-\frac{d\tilde{h}}{d\varphi
}\frac{d\tilde{\chi}}{d\varphi}-\left(  2\frac{d^{2}\tilde{h}}{d\varphi^{2}%
}+2(3\nu-2)\tilde{h}\right)  \tilde{\chi}=-1,\label{A13}%
\end{equation}%
\[
\left.  \frac{d\tilde{\chi}}{d\varphi}\right\vert _{\varphi=0}=0,\qquad
\chi\left(  \frac{\alpha}{2}\right)  =0,
\]
where $\tilde{\chi}$ is defined in the domain $-\alpha/2\leq\varphi\leq
\alpha/2$ and is even in $\varphi$. The boundary problem (\ref{A13}) has a
unique solution, which we plot in Fig.\ \ref{newintegration} for two opening
angles ($\alpha=\pi/4$ and $3\pi/4$).

The introduction of the regularized pressure function $\tilde{\chi}$ is not
special to the uniform evaporation case, and it could be readily defined with
a more general and singular evaporation profile. Function $\tilde{h}$ in
Eq.\ (\ref{A121}) is independent of the evaporation, and its exponent $2$ is
introduced to compensate for the possible singularity of the reduced pressure
function $\tilde{\psi}$ in the uniform evaporation case. With a more general
evaporation rate, singular behavior of $\tilde{\psi}$ may depend on the
singularity of the evaporation rate, and the exponent of $\tilde{h}$ in the
definition (\ref{A121}) should be adjusted accordingly. It can be shown that
in the case of the dry-surface evaporation the exponent becomes $5/2$ due to
the divergence of the evaporation rate at the contact line.

We use both functions $\tilde{\psi}$ and $\tilde{\chi}$ in this paper.
Function $\tilde{\chi}$ is employed to obtain the numerical results because of
its regularity, and function $\tilde{\psi}$ is used in asymptotic analysis
because of its simple asymptotic form and its direct connection to the
physical properties of the system and their singular behavior.

\section{Crossover at the 90-degree opening angle}

It has been shown \cite{5, 8} that in the limit $r\ll R(t)\approx R_{i}$ the
first two leading order terms in the full expansion of the surface height
$h(t,r,\varphi)$ read: for the opening angle $\alpha$ in the vicinity of
$\pi/2$, but not strictly equal to $\pi/2$:
\begin{equation}
h(r,\varphi)=\frac{1}{4}\frac{r^{2}}{R_{i}}\left(  \frac{\cos2\varphi}%
{\cos\alpha}-1\right)  +C\left(  \alpha\right)  \frac{r^{\pi/\alpha}}%
{R_{i}^{\pi/\alpha-1}}\cos\frac{\pi\varphi}{\alpha},\label{111}%
\end{equation}
and at exactly $\alpha=\pi/2$:
\begin{align}
h(r,\varphi)  & =-\frac{1}{\pi}\frac{r^{2}}{R_{i}}\ln\frac{r}{R_{i}}%
\cos2\varphi\nonumber\\
& +\frac{r^{2}}{R_{i}}\left(  \frac{1}{\pi}\varphi\sin2\varphi-\frac{1}%
{4}+C_{0}\cos2\varphi\right)  ,\label{1120}%
\end{align}
where $C\left(  \alpha\right)  $ and $C_{0}$ are related by Eq.\ (\ref{15}),
and $C_{0}$ is independent of $\alpha$. For simplicity, we replaced $R(t)$
with $R_{i}$ and hence suppressed the time dependence of $h$.

As mentioned before, the apparent divergence in Eqs.\ (\ref{14a}),
(\ref{14b}), and hence (\ref{111}) at $\alpha=\pi/2$ is artificial, and, as an
implicit function of $\alpha$, the surface shape $h(r,\varphi)$ is actually
continuous at $\alpha=\pi/2$. This can be readily checked by expanding the
divergent terms in small parameter $(\pi/2-\alpha)$ near $\alpha=\pi/2$ as was
done in Refs.\ 5 and 17. Moreover, as can be shown by the same method, all the
derivatives of $h$ with respect to $r$ and $\varphi$ up to any order are also
continuous at $\alpha=\pi/2$.

We can actually estimate the size of the neighborhood $[\pi/2-\Delta\alpha,
\pi/2+\Delta\alpha]$ where the first and the second leading order terms on the
right side of Eq.\ (\ref{111}) are comparable to each other by comparing the
dominant divergent terms in expressions (\ref{111}) and (\ref{1120}). For
$\alpha<\pi/2$, $\alpha=\pi/2-\Delta\alpha$, we have%
\begin{equation}
\frac{1}{4} \frac{r^{2}}{R_{i}} \frac{\cos2\varphi}{\cos\alpha} \sim- \frac
{1}{\pi} \frac{r^{2}}{R_{i}} \ln\frac{r}{R_{i}} \cos2\varphi,\label{117}%
\end{equation}
or, since $\cos\alpha\approx\Delta\alpha$,
\begin{equation}
\Delta\alpha\sim\frac{\pi}{4} \left\vert \ln\frac{r}{R_{i}} \right\vert
^{-1}.\label{118}%
\end{equation}
For $\alpha>\pi/2$, $\alpha=\pi/2+\Delta\alpha$, and with $C(\alpha)$ given by
Eq.\ (\ref{15}), one can easily obtain:
\begin{equation}
\frac{1}{4\Delta\alpha} \frac{r^{\pi/\alpha}}{R_{i}^{\pi/\alpha- 1}}
\cos2\varphi\sim- \frac{1}{\pi} \frac{r^{2}}{R_{i}} \ln\frac{r}{R_{i}}
\cos2\varphi,\label{119}%
\end{equation}
or, since $(r/R_{i})^{\pi/\alpha- 2} \approx(r/R_{i})^{- 4\Delta\alpha/\pi}$,
\begin{equation}
\Delta\alpha\sim\frac{\pi}{4} \left\vert \ln\frac{r}{R_{i}} \right\vert ^{-1}
\left( \frac{r}{R_{i}}\right) ^{- 4\Delta\alpha/\pi}.\label{120}%
\end{equation}
Since $\Delta\alpha$ is always small whenever $r \ll R_{i}$, the estimates
(\ref{118}) and (\ref{120}) actually provide compatible results.

In asymptotic analysis employed in this paper as well as in Ref.\ 6, only the
term of the smaller exponent of $r$ ($\pi/\alpha\lessgtr2$ according to
$\alpha\gtrless\pi/2$) was retained on the right side of Eq.\ (\ref{111}),
treating $\alpha=\pi/2$ as the limiting case despite a crossover of the
exponents at this value. This approximation does bring about some subtleties
on some occasions, and necessary corrections or modifications of our results
need to be made. Moreover, although the existence of the crossover region
$[\pi/2-\Delta\alpha,\pi/2+\Delta\alpha]$ is uniquely due to the asymptotic
form of surface shape $h$, it does manifest itself in other properties of the
evaporating drop. It is interesting to see how the crossover affects our
asymptotic analysis of the system.

In the case of the velocity field, it was found in Eq.\ (\ref{101}) that
$v_{\varphi}$ at the contact line would vanish in the limit $\alpha
\rightarrow\pi/2$. However, intuitively, according to the conservation of the
fluid mass, locally the velocity should be inversely proportional to the slope
of the surface of the drop. As mentioned above, $\partial h/\partial\varphi$
remains finite and depends continuously on the opening angle, even at
$\alpha=\pi/2$, and therefore $v_{\varphi}|_{\alpha=\pi/2}$ should not be
vanishing as well. Indeed, if, instead of employing Eqs.\ (\ref{14a}) and
(\ref{14b}), we use the leading order term $h(r,\varphi)|_{\alpha=\pi
/2}=-(1/\pi)(r^{2}/R_{i})\ln(r/R_{i})\cos2\varphi$ on the right-hand side of
Eq.\ (\ref{1120}) in Eq.\ (\ref{100}) and let $\nu=2$, we obtain:
\begin{equation}
v_{\varphi}|_{\alpha=\pi/2}\rightarrow\frac{J_{0}}{\rho}\frac{\pi}{2}\left(
\frac{r}{R_{i}}\right)  ^{-1}\left\vert \ln\frac{r}{R_{i}}\right\vert
^{-1},\label{114}%
\end{equation}
which is non-vanishing at the contact line.

Combining Eqs.\ (\ref{101}) and (\ref{114}), we can again identify a crossover
region near the opening angle $\alpha=\pi/2$, which was somewhat concealed by
the asymptotic analysis we employed in this paper as well as in Ref.\ 6. We
believe, as partly shown above, that the velocity field (as well as all other
physical properties of the system) depends on the opening angle $\alpha$
continuously. And in the small neighborhood $[\pi/2-\Delta\alpha, \pi
/2+\Delta\alpha]$ actual physical properties should be interpolated, so that
Eq.\ (\ref{101}), which holds only outside of this neighborhood, is related
continuously to the result (\ref{114}), which applies exactly at $\alpha
=\pi/2$. By comparing Eqs.\ (\ref{101}) and (\ref{114}), we see that
\begin{equation}
\tan\left( \frac{\pi}{2} - \Delta\alpha\right)  \sim\frac{4}{\pi} \left\vert
\ln\frac{r}{R_{i}} \right\vert ,\label{115}%
\end{equation}
and therefore we can find an estimate for $\Delta\alpha$:
\begin{equation}
\Delta\alpha\sim\frac{\pi}{2} - \arctan\frac{4}{\pi} \left\vert \ln\frac
{r}{R_{i}} \right\vert \approx\frac{\pi}{4} \left\vert \ln\frac{r}{R_{i}}
\right\vert ^{-1}.\label{116}%
\end{equation}
Result (\ref{116}) is identical to Eq.\ (\ref{118}). To no surprise, the
crossover originating from the surface was recovered in the result for the
velocity, which was determined using that surface shape. Similar interpolative
estimates can be conducted for all other physical quantities. In principle,
all the results we obtained so far (as well as those in Ref.\ 6) apply only
outside of the neighborhood $[\pi/2-\Delta\alpha, \pi/2+\Delta\alpha]$.

\end{document}